\documentclass{iopart}
\usepackage{graphicx}

\begin{document}

\paper{A one-dimensional model for theoretical analysis of single molecule experiments}
\author{Erik Van der Straeten\footnote{Research Assistant of the Research Foundation - Flanders (FWO - Vlaanderen)} and Jan Naudts}

\address{Departement Fysica, Universiteit Antwerpen, \\Universiteitsplein 1, 2610 Antwerpen, Belgium}
\eads{\mailto{Erik.VanderStraeten@ua.ac.be}, \mailto{Jan.Naudts@ua.ac.be}}

\begin{abstract}
In this paper we compare two polymer stretching experiments. The outcome of both experiments is a force-extension relation. We use a one-dimensional model to show that in general the two quantities are not equal. In certain limits, however, both force-extension relations coincide.
\end{abstract}

\section{Introduction}
Traditional experiments on biological molecules are carried out on large groups of molecules. Such groups of molecules are macroscopic thermodynamic systems which can be interpreted using conventional thermodynamics. Recently, it became possible to experimentally manipulate individual molecules. Such single molecules are investigated for several reasons, some of which are mentioned below.

The mechanical properties of single molecules are important in processes like polymer folding \cite{ref4,ref9} and DNA strand separation \cite{ref5}.

Biophysicists study small motors that are responsible for converting chemical energy to useful forms of work. These biological machines work under non-equilibrium conditions. The working principles are poorly understood. Many biomolecules can function as such motors. So the study of single molecules can help to understand the properties of biological machines \cite{ref6,ref7}.

The experiments \cite{ref1,ref8} of interest for the present paper combine results obtained under varying thermodynamic circumstances. In one type of experiment, the two ends of a single linear molecule are held at fixed positions and the exerted forces are measured. In the other type of experiment, the force exerted on the ends is constant and the position of the end points is measured. The corresponding ensembles differ because in one case the end positions are fluctuating quantities while they are kept constant in the other case. Nevertheless, the outcomes of the two experiments should be related, as argued in \cite{ref1}. In general, results of statistical physics do not depend on the choice of ensemble. This equivalence of ensembles has been proven for large classes of systems. However, it may fail in case of single molecule experiments for at least two reasons. The proof of ensemble equivalence relies on the thermodynamic limit, while here the number of degrees of freedom is limited. It may also break down because long range interactions, in particular those due to excluded volume, are important. The present paper is based on the model studied in \cite {ref2}. This model is simple enough so that statements can be made which do not depend on the assumption of a large number of degrees of freedom.

In the next section we discuss the ensemble dependence of measurements on single molecules in more detail. In section \ref{model} we give a short review of the results obtained in \cite{ref2} and we show how this results can be applied on single molecules. In sections \ref{isotensial} and \ref{isometric} we derive expressions for the force-extension relation for two different ensembles, the canonical and the grand canonical ensemble. In section \ref{compa} we compare the force-extension relations of both experiments. The final section gives a short discussion of the results.

\section{Single molecule experiments}
In the next two paragraphs we discuss two types of experiment, first from experimental and then from theoretical point of view.

The first experiment is the ideal isotensional experiment where the forces on the end points of the molecule are kept constant and the fluctuating end-to-end distance is measured. The second experiment is the ideal isometric experiment where the end points of the molecule are held fixed and the fluctuating force is measured. A clear description of a possible experimental setup can be found in \cite{ref3}. The outcome of both experiments is a force-extension curve. In the isotensional experiment, one obtains the average end-to-end distance $\langle x\rangle(F)$ as a function of the applied force $F$. In the isometric experiment, one obtains the average force $\langle F\rangle(x)$ as a function of the fixed end-to-end distance $x$. Inverting $\langle x\rangle(F)$ results in $F(\langle x\rangle)$. An interesting question is whether $\langle F\rangle(x)$ and $F(\langle x\rangle)$ should coincide.

From a thermodynamic point of view, the end-to-end distance $x$ is a (non-conserved) observable and the force $F$ is the corresponding control parameter (just like temperature is the control parameter of the energy). By measurement of the average value of the observables one can estimate the value of the corresponding control parameters. In this view, the canonical ensemble is a one parameter family. Such family has only one fluctuating observable (mostly the energy) and one control parameter (mostly temperature). Hence, the isometric experiment is described by a canonical ensemble because only the energy fluctuates. The isotensional experiment requires a grand canonical ensemble because both the energy and the end-to-end distance fluctuate. One can derive expressions for $\langle F\rangle(x)$ and $F(\langle x\rangle)$ in both ensembles. Whether the expressions are equal or not, boils down to a question of equivalence of a canonical and a grand canonical ensemble.

In \cite{ref1} Keller \etal investigate the ensemble dependence of single molecule measurements. They replace the familiar thermodynamic potentials (Gibbs or Helmholtz free energy) by appropriate potentials of mean force. Keller \etal find that the two experiments should be related by a Laplace transform. In the thermodynamic limit (infinite long chain) the Laplace transform is dominated by it's saddle point value and the two experiments are then related by a Legendre transform \cite{ref8}. Following this reasoning, the force-extension relations of both experiments coincide. However, in general the saddle point approximation is not valid. Therefore the experimentally obtained force-extension relations are distinct.

\section{Model} \label{model}
In \cite{ref2} we have studied a non-Markovian random walk in one dimension. It depends on two parameters $\epsilon$ and $\mu$, the probabilities to go straight on when walking to the right, respectively to the left. We have calculated the joint probability distribution $P_n(x,k,\epsilon,\mu)$. This is the probability to end in position $x$ with $k$ reversals of direction after $n$ steps. The random walk can be used as a simple model of a polymer with $n$ atoms. Energy $H$ is proportional to the number of kinks $k$ and is given by $H=-hk$. This means that the polymer is totally folded at zero temperature. The position of the end point $x$ measures the effect of an external force applied to the end point.

In \cite{ref2} we have calculated an exact expression for $P_n(x,k,\epsilon,\mu)$. It is the product of two contributions
\begin{equation}
P_n(x,k,\epsilon,\mu)=C_n(x,k)F_n(x,k,\epsilon,\mu).
\end{equation}
The function $C_n(x,k)$ counts the number of walks which start in the origin and end in $x$ after $n$ steps with $k$ kinks. This function only depends on the final values of $x$ and $k$. The probability that an $n$-step walk ends in $x$ and has $k$ kinks is denoted $F_n(x,k,\epsilon,\mu)$. This probability depends on the values of $\epsilon$, $\mu$ and the initial conditions. See figure \ref{example:prob} for an example of a walk with $6$ steps and $4$ kinks, ending in $x=0$ and starting to the right. There are two possible configurations $C_6(0,4)=2$. The probability for both configurations is $F_6(0,4,\epsilon,\mu)=p_0(1-\epsilon)^2(1-\mu)^2\mu$.
\setlength{\unitlength}{1mm}
\begin{figure}
\begin{picture}(20,55)
\put(45,13){$p_0$}\put(43,19){$1-\epsilon$}\put(43,19){$1-\epsilon$}\put(25,23){$\mu$}\put(22,32){$1-\mu$}\put(22,41){$1-\epsilon$}\put(22,50){$1-\mu$}
\qbezier(35,10)(75,15)(35,20)\qbezier(35,20)(-5,25)(25,30)\qbezier(25,30)(45,35)(25,40)\qbezier(25,40)(-5,45)(35,50)
\put(101,13){$p_0$}\put(98,21){$1-\epsilon$}\put(98,32){$1-\mu$}\put(98,40){$1-\epsilon$}\put(80,43){$\mu$}\put(78,50){$1-\mu$}\put(22,50){$1-\mu$}
\qbezier(90,10)(130,15)(100,20)\qbezier(100,20)(80,25)(100,30)\qbezier(100,30)(130,35)(90,40)\qbezier(90,40)(50,45)(90,50)
\put(10,0){\line(1,0){50}}\put(70,0){\line(1,0){45}}
\put(35,1){0}\put(90,1){0}\put(55,1){2}\put(112,1){2}\put(13,1){-2}\put(72,1){-2}
\put(35,10){\circle*{1}}\put(55,15){\circle*{1}}\put(35,20){\circle*{1}}
\put(12.5,26){\circle*{1}}\put(35,35){\circle*{1}}\put(12.5,44){\circle*{1}}\put(35,50){\circle*{1}}
\put(90,10){\circle*{1}}\put(112.5,15.5){\circle*{1}}\put(90,25){\circle*{1}}
\put(112.5,34.5){\circle*{1}}\put(90,40){\circle*{1}}\put(70,45){\circle*{1}}\put(90,50){\circle*{1}}
\end{picture}
\caption{\label{example:prob}Two walks with $6$ steps which end in $x=0$, have the first step to the right and have $4$ kinks. $\epsilon$ and $\mu$ are the probabilities to go straight on when walking to the right, respectively to the left. $p_0$ is the probability that the first step goes to the right.}
\end{figure}
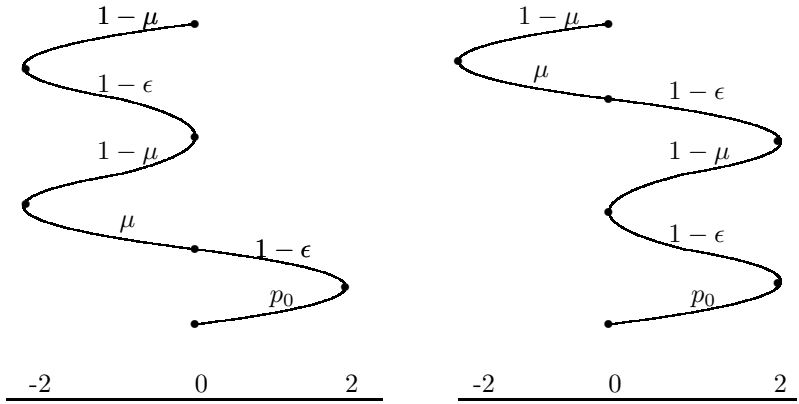

One has approximately
\begin{equation}
P_n(x,k,\epsilon,\mu)\approx p_n(x,k,\epsilon,\mu)=c_n(x,k)f_n(x,k,\epsilon,\mu)
\end{equation}
\begin{equation}
f_n(x,k,\epsilon,\mu)=(1-\epsilon)^{\frac{k}{2}}(1-\mu)^{\frac{k}{2}}\epsilon^{\frac{(n-k)a+x}{2a}}\mu^{\frac{(n-k)a-x}{2a}}
\end{equation}
\begin{equation}
c_n(x,k)=\frac{\Gamma\left(\frac{na-x}{2a}+1\right)\Gamma\left(\frac{na+x}{2a}+1\right)}
{\Gamma\left(\frac{k}{2}+1\right)^2\Gamma\left(\frac{(n-k)a-x}{2a}+1\right)\Gamma\left(\frac{(n-k)a+x}{2a}+1\right)},
\end{equation}
The lattice parameter is denoted $a$. The approximate distribution $p_n(x,k,\epsilon,\mu)$ deviates from the exact distribution $P_n(x,k,\epsilon,\mu)$ for two reasons. Firstly, $P_n(x,k,\epsilon,\mu)$ has a small dependence on the initial conditions. Secondly, the expression $P_n(x,k,\epsilon,\mu)$ is different for even or odd $k$. These effects are negligible when the number of atoms $n$ is large. In the present paper, we only use the approximate distribution $p_n(x,k,\epsilon,\mu)$.

Define the following reparametrisation
\begin{equation}
F=\frac{1}{2a\beta}\ln\frac{\epsilon}{\mu} \textrm{\ \ and\ \ }\beta=\frac{1}{2h}\ln\frac{(1-\epsilon)(1-\mu)}{\epsilon\mu}.
\label {Fparam}
\end{equation} 
Then the probability $p_n(x,k)$ (omitting the dependence on $\epsilon$ and $\mu$) can be written as
\begin{equation} \label{probgroot}
p_n(x,k)=\frac{c_n(x,k)}{Z_n} e^{-\beta(-hk-Fx)}.
\end{equation}
with the partition sum $Z_n$ approximately given by
\begin {equation}
Z_n\simeq(\epsilon\mu)^{-n/2}.
\end {equation}
In this form, it is clear that $p_n(x,k)$ is a Gibbs distribution. We will show that the parameter $\beta$ is the inverse temperature and $F$ is the applied external force.

The model depends on two parameters, the probabilities $\epsilon$ and $\mu$ (or the inverse temperature $\beta$ and the applied external force $F$). Hence measurement of two observables is needed to estimate the values of all parameters of the model. In \cite{ref1} we obtained exact expressions for the average number of kinks and the average end-position
\begin{equation} \label{avgroot}
\langle k\rangle=2n\frac{(1-\epsilon)(1-\mu)}{2-\epsilon-\mu} \textrm{\ \ ,\ \ } \langle x\rangle=na\frac{\epsilon-\mu}{2-\epsilon-\mu}.
\end{equation}
These relations link the model parameters $\epsilon$ and $\mu$ to the measurable quantities $\langle x\rangle$ and $\langle k\rangle$.

\section{Isotensial experiment} \label{isotensial}
The isotensial experiment is described in the grand canonical ensemble, with the end-to-end distance $x$ and the energy $H=-hk$ both fluctuating. The probability for a polymer with $n+1$ atoms to end in $x$ with $k$ kinks is $p_n(x,k)$ and is given by (\ref{probgroot}). The entropy is
\begin{equation}\label{grootentr}
\frac 1{k_{\rm B}} S=-\sum_k\sum_x p_n(x,k)\ln\frac{p_n(x,k)}{c_n(x,k)}=-\beta h\langle k\rangle-\beta F\langle x\rangle+\ln Z_n.
\end{equation}
Hence, the free energy $G$ equals
\begin {eqnarray}
 G&=&E-F\langle x\rangle-TS=-\beta^{-1}\ln Z_n.
\end {eqnarray}
From $1=\sum_{x,k}p_n(x,k)$ and (\ref {probgroot}) then follows
\begin {eqnarray}
0&=&\frac {\partial\,}{\partial\beta}\sum_{x,k}p_n(x,k)
=-E+F\langle x\rangle-\frac {\partial\,}{\partial\beta}\ln Z_n\\
0&=&\frac {\partial\,}{\partial F}\sum_{x,k}p_n(x,k)
=\beta\langle x\rangle-\frac{\partial\,}{\partial F}\ln Z_n.
\end {eqnarray}
With these relations one shows that the free energy $G$ satisfies the relations
\begin {eqnarray}
\frac {\partial\,}{\partial\beta}\beta G&=&E-F\langle x\rangle\\
\frac {\partial\,}{\partial F}G&=&-\langle x\rangle.
\end {eqnarray}
This proves that $k_{\rm B}\beta$ is the inverse of the thermodynamic temperature $T$ and that $F$ is the applied force.
Note that $T$ should be positive. This gives the condition $\epsilon+\mu\leq1$.

\begin{figure}
 \begin{center}
 \includegraphics[width=0.7\textwidth]{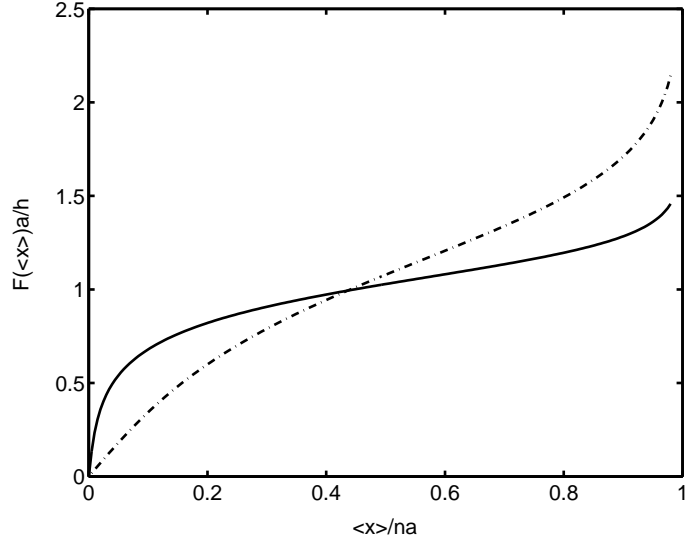}
 \end{center}
 \caption{Plot of $F(\langle x\rangle)a/h$ as a function of $\langle x\rangle/na$. The value of $\beta h$ is $5$ for the solid line and $2$ for the dotted line.}
 \label{fig:Fgroot}
\end{figure}

By inverting (\ref{avgroot}), one gets expressions for $\epsilon$ and $\mu$ as a function of $\langle k\rangle$ and $\langle x\rangle$
\begin {eqnarray}
\epsilon&=&\frac {a(n-\langle k\rangle)+\langle x\rangle}{na+\langle x\rangle}\\
\mu&=&\frac {a(n-\langle k\rangle)-\langle x\rangle}{na-\langle x\rangle}.
\end {eqnarray}
In combination with (\ref {Fparam}) this gives
\begin {equation} \label{Fbetagroot}
F
=\frac{1}{2a\beta}\ln\left(\frac{(n-\langle k\rangle)a+\langle x\rangle}{na+\langle x\rangle}\frac{na-\langle x\rangle}{(n-\langle k\rangle)a-\langle x\rangle}\right).
\end {equation}
In the latter expression $\langle k\rangle$ can be eliminated using
\begin{eqnarray}
\langle k\rangle&=&\frac{1}{1-e^{-2\beta h}}\left(n-\frac 1a\sqrt{\langle x\rangle^2\left(1-e^{-2\beta h}\right)+n^2a^2e^{-2\beta h}}\right).
\end{eqnarray}
The result is $F$ as a function of $\langle x\rangle$ at constant temperature.
It is plotted in figure \ref{fig:Fgroot} for two values of $\beta h$.

\section{Isometric experiment} \label{isometric}
We proceed with the isometric experiment, which is described in the canonical ensemble, with fixed end-to-end distance $x$. Without loss of generality, we assume that $x\ge 0$. The probability distribution is the conditional distribution obtained from $p_n(x,k)$ by requiring that $x$ has a given value
\begin{equation} \label{condprob}
p_n(k|x)=\frac{p_n(x,k)}{\sum_kp_n(x,k)}=\frac{c_n(x,k)e^{\beta h k}}{\sum_{k}c_n(x,k)e^{\beta hk}}=\frac{1}{z_n}c_n(x,k)e^{\beta hk},
\end{equation}
with $z_n=\sum_{k}c_n(x,k)e^{\beta hk}$ the canonical partition function. The entropy is
\begin{equation} \label{Skanex}
\frac 1{k_{\rm B}}S=-\sum_k p_n(k|x)\ln\frac{p_n(k|x)}{c_n(x,k)}=-\beta h\langle k\rangle+\ln z_n.
\end{equation}
The free energy is
\begin {equation}
G=E-TS=-\beta^{-1}\ln z_n.
\end {equation}
The mean force can be obtained by taking the derivative of $G$ with respect to $x$. The latter is a discrete variable, so the force becomes
\begin{equation} \label{Fkanex}
\langle F\rangle(x)=\frac{\Delta G}{\Delta x}=-\frac{1}{\beta}\frac{\Delta\ln z_n}{\Delta x}.
\end{equation}
We cannot calculate this discrete derivative in closed form. In the next section we proceed by assuming that $k$ and $x$ are continuous variables. Under this assumption an explicit expression for $\langle F\rangle(x)$ can be obtained. The differences between this expression and expression (\ref {Fkanex}) will be discussed in detail. See also Fig.\,\ref {fig:Fcondis}A.

\subsection{Continuum approximation}
Assuming that $k$ and $x$ are continuous variables ($n$, $k$ and $x$ large), one can replace summations by integrations. In \ref{app1} we derive under this assumption following expression for the force
\begin{equation} \label{Fkanco}
\langle F\rangle(x)
=-\frac{1}{\beta}\left\langle \frac{1}{c_n(x,k)}\frac{\partial c_n(x,k)}{\partial x}\right\rangle
+\frac{1}{\beta}\frac{1}{a}p_n(n-x/a|x).
\end{equation}
In the continuum approximation, the force consists of two contributions. The first contribution is due to the change of entropy, the second is a finite size effect (this term disappears in the limit $n\rightarrow\infty$). 
\begin{figure}
\parbox{6cm}{\includegraphics[width=6cm]{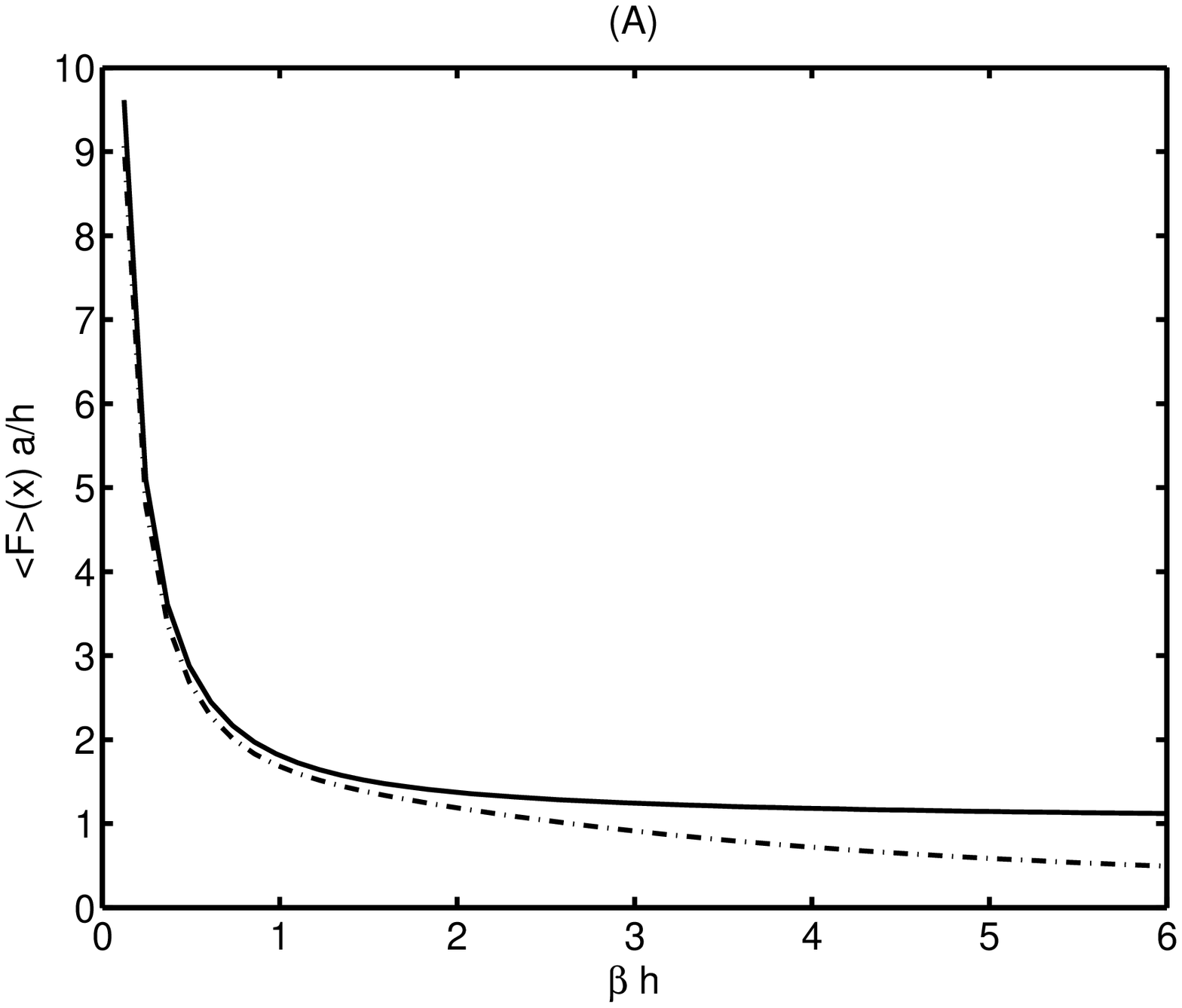}}
\hfill
\parbox{6cm}{\includegraphics[width=6cm]{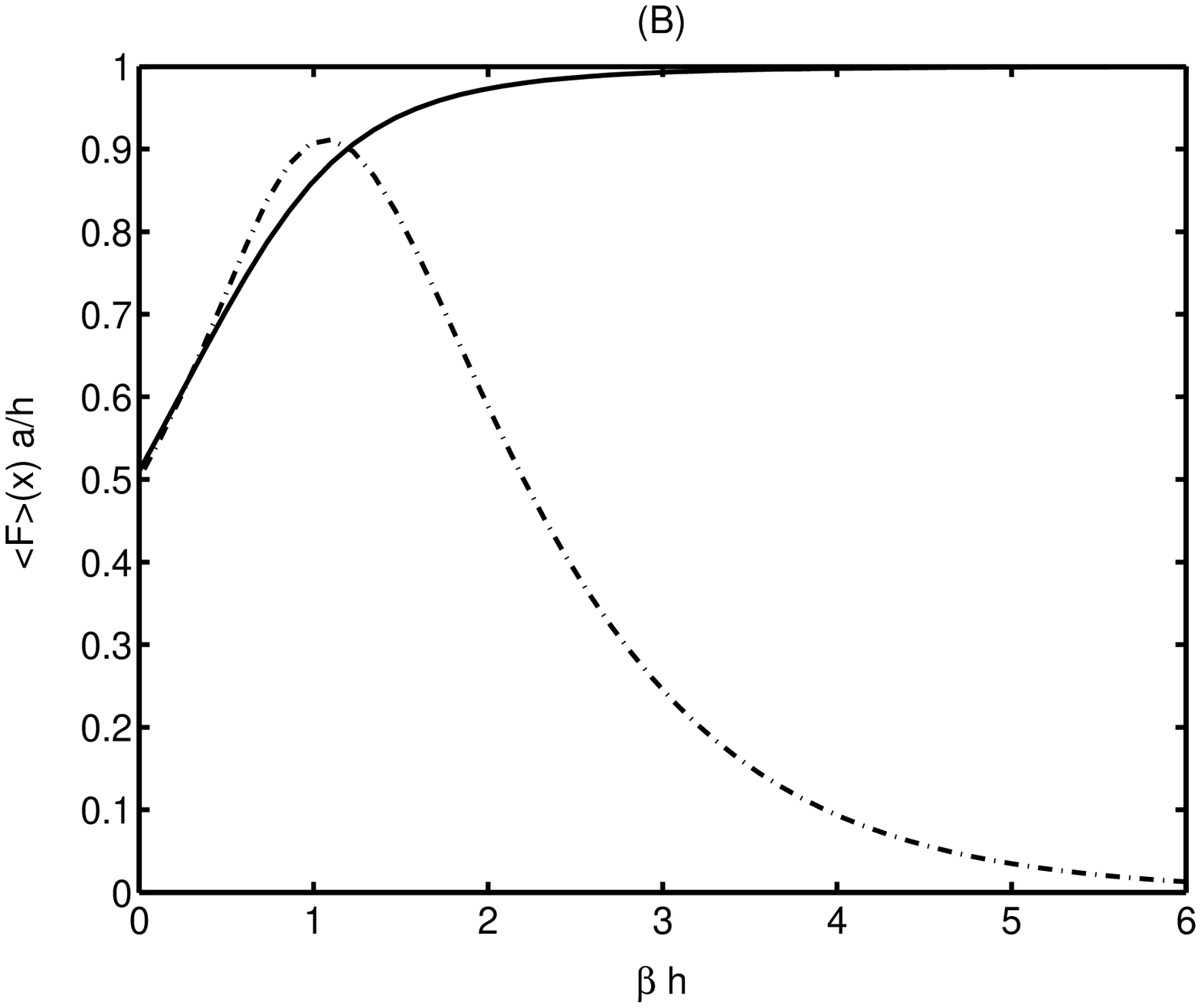}}
\caption{\label{fig:Fcondis} (A) Plot of the isometric force as a function of $\beta h$, with $n=150$ and $x=na/2$. An exact expression for this force is given by (\ref{Fkanex}) (solid line). After the continuum approximation we obtain (\ref{Fkanco}) (dotted line). (B) Plot of the adiabatic force as a function of $\beta h$, with $n=150$ and $x=na/2$. An exact expression for this force is given by (\ref{isoentropic_Fdis}) (solid line). After the continuum approximation we obtain (\ref{isoentropic_Fcon}) (dotted line). }
\end{figure} 

Expressions (\ref{Fkanex}) and (\ref{Fkanco}) differ in the limit $\beta\rightarrow\infty$. Indeed, the limiting value of expression (\ref{Fkanex}) is
\begin{equation} \label{dfff}
\lim_{\beta\rightarrow\infty}\langle F\rangle(x)=-\frac{\Delta}{\Delta x}\lim_{\beta\rightarrow\infty}\langle hk\rangle=-h\frac{\Delta}{\Delta x}\left(n-\frac{x}{a}\right)=\frac{h}{a}.
\end{equation}
The limiting value of expression (\ref{Fkanco}) is $0$. The force vanishes in the limit $n\rightarrow\infty$ because an infinite chain implies that the average number of kinks is infinite, independently of the end-to-end distance. This has the immediate consequence that for low temperatures the force equals zero. 

Mathematically, the origin of the difference between the force for a finite or an infinite chain, is the order of taking the limit $n\rightarrow\infty$ and performing the derivative with respect to $x$. This can be seen by taking the limit $n\rightarrow\infty$ of expression (\ref{dfff})
\begin{equation}
-\lim_{n\rightarrow\infty}\frac{\Delta}{\Delta x}\lim_{\beta\rightarrow\infty}\langle hk\rangle=\frac{h}{a}\textrm{\ \ \ }\neq\textrm{\ \ \ } -\frac{\partial}{\partial x}\lim_{n\rightarrow\infty}\lim_{\beta\rightarrow\infty}\langle hk\rangle=0.
\end{equation}
We conclude that the order of taking limits influences the result for the force. To illustrate this we plotted expressions (\ref{Fkanex}) and (\ref{Fkanco}) as functions of $\beta$ (with $n$ and $x$ large) in Figure \ref{fig:Fcondis}A. The figure demonstrates that the continuum approximation holds for small values of $\beta$, but not for large values of $\beta$, because of the different limiting values.

\subsection{Adiabatic experiment}
If the experiment is carried out fast enough, the entropy is constant and does not contribute to the adiabatic force, which is given by $\langle F\rangle=-\Delta\langle k\rangle/\Delta x$. The results obtained in this case are found in \ref{app2}. Figure \ref{fig:Fcondis}B shows expressions (\ref{isoentropic_Fdis}) and (\ref{isoentropic_Fcon}) as functions of $\beta$. Again, the continuum approximation holds for small values of $\beta$, but not for large values of $\beta$. Figure \ref{fig:Fcondis}B shows also that the discrete force is an increasing function of $\beta$. We can understand this by following reasoning.

In the limit $\beta\rightarrow\infty$, the isometric experiment and the adiabatic experiment coincide, because in this limit the free energy is equal to the energy. The average number of kinks equals the maximum that is allowed under the constraint of the fixed end-to-end distance $x$. Extending the chain with $\Delta x$, results in a decrease of the average number of kinks by $\Delta x/a$. So the force becomes $h/a$. For lower values of $\beta$ the average number of kinks decreases. Then an extension of the chain by $\Delta x$ can still be compensated by a decrease of the number of kinks, but also by a rearrangement of the kinks, without changing their number. So the decrease of the average number of kinks is smaller than $\Delta x/a$ and the force is smaller than $h/a$.

The difference between the isometric and the adiabatic force can most clearly be seen in the limit $\beta\rightarrow0$. Figures \ref{fig:Fcondis}A and \ref{fig:Fcondis}B demonstrate that the isometric force diverges in the high temperature limit, while the adiabatic force converges to a constant. This limiting behaviour can be calculated in closed form using the original, non-approximated expressions for $c_n(x,k)$ as defined in \cite{ref2}. The high temperature limit of the number of kinks is
\begin{equation}
\lim_{\beta\rightarrow 0}\langle k\rangle=\frac{n}{2}\left[1-\left(\frac{x}{na}\right)^2\right].
\end {equation}
The high temperature limit of the adiabatic force is
\begin{equation} \label{limFisoe}
\lim_{\beta\rightarrow 0}\langle F\rangle(x)=\frac{h}{a}\frac{x}{na},
\end{equation}
while the limiting value of the isometric force is
\begin{equation} \label{limFisot}
\langle F\rangle(x)\sim-\frac 1\beta\frac{\Delta}{\Delta x}\ln \sum_kc_n(x,k)\approx\frac 1{\beta a}\ln\frac{1+x/na}{1-x/na}.
\end{equation}
Stirling's approximation and the continuum approximation have been used to obtain the last expression.

\section{Comparison of the two force-extension relations} \label{compa}
\begin{figure}
\parbox{6cm}{\includegraphics[width=6cm]{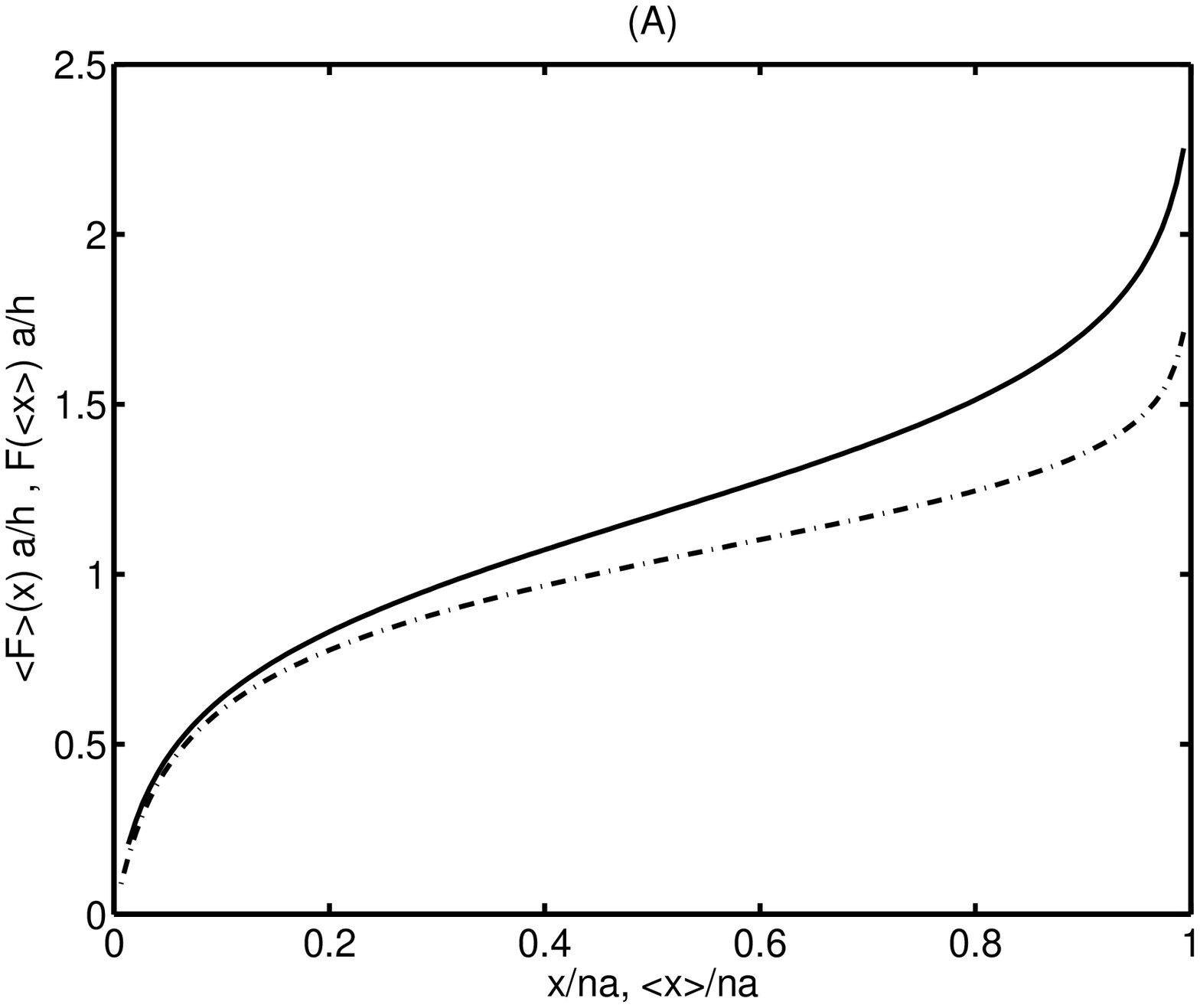}}
\hfill
\parbox{6cm}{\includegraphics[width=6cm]{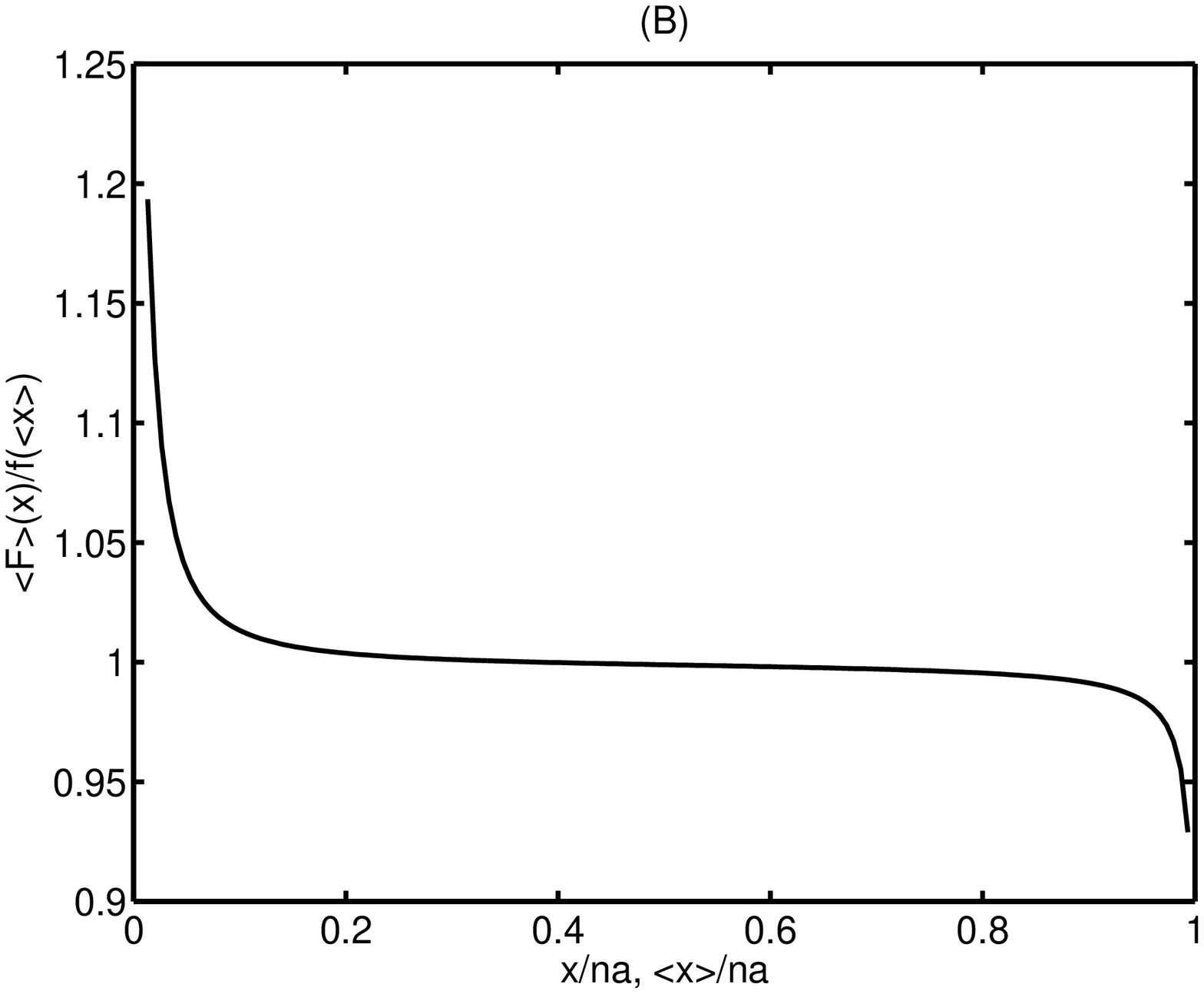}}
\caption{\label{fig:Fgrootkano} (A) Plot of the grand canonical force (dotted line) as a function of average end-to-end distance and the canonical force (solid) line as a function of end-to-end distance (with $n=150$ and $\beta h=4$). (B) Plot of the quotient $\langle F\rangle(x)/f(\langle x\rangle)$ as a function of (average) end-to-end distance (with $n=150$ and $\beta h=4$) --- See (\ref {fdef}).}
\end{figure}

Figure \ref{fig:Fgrootkano}A shows the exact force-extension relations for the grand canonical (\ref{Fbetagroot}) and canonical (\ref{Fkanex}) ensembles. As can be seen, they differ. In certain limits however the two force-extension relations do coincide.

\subsection{Thermodynamic limit of the force-extension relations}
Assume that $\langle x\rangle/na\approx 1/2$ and $\langle k\rangle/n\approx 1/2$. Then one can approximate the grand canonical force by
\begin{equation}
F(\langle x\rangle)\approx f(\langle x\rangle)-\frac{1}{a\beta}\frac{\langle x\rangle}{na}.
\end{equation}
with $f(x)$ defined by
\begin {equation}\label {fdef}
f(x)=\frac{1}{2a\beta}\ln\left(\frac{(n-\langle k\rangle)a+x}{(n-\langle k\rangle)a-x}\right)
\end {equation}
Figure \ref{fig:Fgrootkano}B shows the quotient $\langle F\rangle(x)/f(\langle x\rangle)$. From this figure it is obvious that the canonical force and $f(\langle x\rangle)$ are approximately the same for most values of $\langle x\rangle/na$. The grand canonical force equals $f(\langle x\rangle)$ up to a term proportional to $\langle x\rangle/na$. So the force-extension relations of the grand canonical ensemble and the canonical ensemble are equivalent if $\langle x\rangle/na<<f(\langle x\rangle)$. This inequality holds when $\langle x\rangle/na\approx 1/2$ and $\langle k\rangle/n\approx 1/2$.

\subsection{Physical explanation}
In the grand canonical ensemble, one can calculate the average value $\langle x\rangle(F)$ of the end-to-end distance for any given temperature $T$ and external force $F$, using the probability distribution $p_n(x,k)$. In general, this average value does not coincide with the most probable value $x_{\rm mp}$ of the end-to-end distance. In \cite{ref1}, Keller \etal argue that the difference between the two experiments is a consequence of the difference between $\langle x\rangle(F)$ and $x_{\rm mp}$. We will follow the lines of \cite{ref1} to prove this statement for our model.

First consider the isotensial experiment. With a fixed force and a given temperature one can calculate $\langle x\rangle(F)$. After inverting this relation one obtains $F(\langle x\rangle)$. The most probable value of the end-to-end distance can be calculated by solving the equation
\begin{equation} \label{pogr}
\left.\frac{\Delta}{\Delta x}\sum_kp_n(x,k)\right|_{x=x_{\rm mp}}=0.
\end{equation}
Based on numerical evidence, we assume that this equation has only one solution.

Next consider the isometric experiment, with fixed end-to-end distance taken equal to $x_{\rm mp}$. Using formula (\ref{Fkanex}) one obtains for the average force
\begin{equation} \label{vglkl2}
\langle F\rangle(x_{\rm mp})=-\frac{1}{\beta}\left.\frac{\Delta}{\Delta x}\ln\sum_{k/h=0}^{n-x/a}c_n(x,k)e^{\beta hk}\right|_{x=x_{\rm mp}}.
\end{equation}
In the continuum approximation following equation holds
\begin{equation} \label{vwd}
F\left(\langle x\rangle\right)=\langle F\rangle(x_{\rm mp}).
\end{equation}
This is the same result as that obtained by Keller \etal  \cite{ref1}. It means that the average force $\langle F\rangle$ measured in an isometric experiment equals to the fixed force $F$ of the isotential experiment provided that the fixed extension of the isometric experiment is $x_{\rm mp}$ instead of the average end-to-end distance of the isotential experiment $\langle x\rangle$. Therefore, the force-extension relations of the isometric and isotential experiment coincide only when $\langle x\rangle=x_{\rm mp}$. This is the case in the limit of large $n$, when fluctuations are negligible.

\section{Discussion}
In this paper we study two polymer stretching experiments. The theoretical description involves two different ensembles, one canonical, the other grand canonical. We have derived expressions for the force-extension relations in both ensembles using a simple model.

The model is a one-dimensional random walk depending on two parameters. We showed in \cite{ref2} that the probability distribution of the walk is approximately a Gibbs distribution, see (\ref{probgroot}). In the present paper we ignore the small deviations from the Gibbs distribution. The model represents a polymer which has $k$ kinks and end-to-end distance $x$. The averages $\langle x\rangle$ and $\langle k\rangle$, or the corresponding control parameters external force $F$ and temperature $T$, are used to determine the two model parameters.

In the grand-canonical description the two observables $k$ and $x$ both fluctuate. The external force $F$ is modelled as usual by adding a term $-Fx$ to the Hamiltonian. The force $F(\langle x\rangle)$ as function of the average end-to-end distance is obtained by inversion of $\langle x\rangle$ as a function of $F$.
In the canonical ensemble, the end-to-end distance $x$ is fixed. The probability distribution for this experiment is obtained from the grand canonical distribution by conditioning on $x$. The average force $\langle F\rangle$ is then obtained as minus the gradient of the free energy
(or of the energy in the adiabatic case). In this way one obtains the average force $\langle F\rangle(x)$ as function of the fixed end-to-end distance $x$. A peculiarity is that the result depends slightly on the order of taking the continuum limit and calculating the gradient.

In general the two force-extension relations $F(\langle x\rangle)$ and $\langle F\rangle(x)$ are different, which can clearly be seen in figure \ref{fig:Fgrootkano}A. In \cite{ref1}, Keller \etal argue that the difference between the experiments is due to the difference between the average end-to-end distance calculated in the grand canonical ensemble and the most probable end-to-end distance of the grand canonical distribution. Our calculations confirm this point of view. But we find additional effects due to finiteness of the polymer chain. All these effects disappear in the limit of infinite long chains, so that the two experiments coincide in the thermodynamic limit.

It is highly desirable to extend our work to two or three dimensions because the role of entropy is too limited in $d=1$. This is the reason why the present model does not exhibit a folding transition.

\appendix
\section{Continuum limit} \label{app1}
Assume that $x$ and $k$ are continuous variables. Then one can write
\begin{equation}
\langle F\rangle(x)=-\frac{1}{\beta}\frac{\partial}{\partial x}\ln z_n=-\frac{1}{\beta}\frac{1}{z_n}\frac{\partial z_n}{\partial x}.
\end{equation}
The partition function becomes
\begin{equation}
\fl z_n=\int_{0}^{n-x/a}{\rm d}k\, c_n(x,k)e^{\beta k}=\int_{0}^{\infty}{\rm d}k\, c_n(x,k)e^{\beta hk}\Theta(n-x/a-k)
\end{equation}
with $\Theta(x)=0$ if $x<0$ and $\Theta(x)=1$ if $x>0$. Calculating the derivative gives
\begin{eqnarray}
\fl \langle F\rangle(x)&=&\frac{1}{\beta}\frac{1}{z_n}\left(\frac{1}{a}\int_{0}^{\infty}{\rm d}k\, c_n(x,k)e^{\beta hk}\delta(n-x/a-k)-\int_{0}^{n-x/a}{\rm d}k\, \frac{\partial c_n(x,k)}{\partial x}e^{\beta hk}\right)
\cr
\fl&=&\frac{1}{\beta}\frac{1}{a}p_n(n-x/a|x)-\frac{1}{\beta}\left\langle \frac{1}{c_n(x,k)}\frac{\partial c_n(x,k)}{\partial x}\right\rangle
\end{eqnarray}

\section{Adiabatic experiment} \label{app2}
If entropy is constant, force is obtained by taking the derivative of the energy
\begin{equation} \label{isoentropic_Fdis}
\langle F\rangle(x)=\frac{\Delta E}{\Delta x}=A(x)+B(x),
\end{equation}
with
\begin{eqnarray} \label{AandB}
\fl A(x)=\frac{\sum_{k=0}^{n-x/a} c_n(x,k)e^{\beta hk}}{\sum_{k=0}^{n-x/a-1} c_n(x+a,k)e^{\beta hk}}\times
\frac{\sum_{k=0}^{n-x/a} c_n(x,k)e^{\beta hk}}{\sum_{k=0}^{n-x/a+1} c_n(x-a,k)e^{\beta hk}}\times
\cr
\fl\Bigg[\left\langle hk\frac{\left\lfloor c_n(x,k)\right\rfloor}{c_n(x,k)}\right\rangle
\left\langle  \frac{1}{c_n(x,k)}\frac{\Delta c_n(x,k)}{\Delta x}\right\rangle-\left\langle \frac{hk}{c_n(x,k)}\frac{\Delta c_n(x,k)}{\Delta x}\right\rangle
\left\langle \frac{\left\lfloor c_n(x,k)\right\rfloor}{c_n(x,k)}\right\rangle\Bigg]
\cr
\fl B(x)=\frac{h}{2a}p_n(n-x/a+1|x-a)\left[n-\frac{x}{a}+1-\frac{\sum_{k/h=0}^{n-x/a}kc_n(x+a,k)e^{\beta hk}}{\sum_{k/h=0}^{n-x/a-1} c_n(x+a,k)e^{\beta hk}}\right]
\cr
\fl+\frac{h}{2a}p_n(n-x/a|x+a)
\left[n-\frac{x}{a}-\frac{\sum_{k=0}^{n-x/a}kc_n(x-a,k)e^{\beta hk}}{\sum_{k=0}^{n-x/a+1} c_n(x-a,k)e^{\beta hk}}\right],
\end{eqnarray}
where we used the notation
\begin{equation}
\left\lfloor c_n(x,k)\right\rfloor=\frac{1}{2}\left[c_n(x+a,k)+c_n(x-a,k)\right].
\end{equation}
In the continuum approximation the force is given by
\begin{eqnarray} \label{isoentropic_Fcon}
\langle F\rangle(x)
&=&-\left\langle hk\frac{\partial\ln c_n(x,k)}{\partial x}\right\rangle
+\left\langle hk\right\rangle
\left\langle \frac{\partial\ln c_n(x,k)}{\partial x}\right\rangle\cr
& &
+\frac{h}{a}\left(n-\frac{x}{a}
-\langle k\rangle\right)p_n(n-x/a|x).
\end{eqnarray}

\end{document}